\documentclass{elsart5p}
\usepackage{amssymb}
\usepackage{longtable}
\usepackage{graphicx}  % standard LaTeX graphics tool
\begin{document}
\renewcommand{\thefootnote}{\fnsymbol{footnote}}

\begin{frontmatter}


\title{Evaluation of fusion-evaporation cross-section calculations}

\author[label1,label2]{B. Blank}\footnotemark[1],
%\ead{blank@cenbg.in2p3.fr}
\author[label1]{G. Canchel},
\author[label1]{F. Seis}\footnotemark[2],
\author[label3]{P. Delahaye}
\address[label1]{Centre d'Etudes Nucl\'eaires de Bordeaux Gradignan, 19 Chemin du Solarium, CS10120, F-33175 Gradignan Cedex, France }
\address[label2]{ISOLDE/CERN, EP Department, CH-1211 Geneve 23, Switzerland}
\address[label3]{Grand Acc{\'e}l{\'e}rateur National d'Ions Lourds, Bd Henri Becquerel, BP 55027, \\14076 CAEN Cedex 05, France}

\begin{abstract}
Calculated fusion-evaporation cross sections from five different codes are compared to experimental data. The present
comparison extents over a large range of nuclei and isotopic chains to investigate the evolution of experimental and calculated
cross sections. All models more or less overestimate
the experimental cross sections. We found reasonable agreement by using the geometrical average of the five model calculations
and dividing the average by a factor of 11.2. More refined analyses are made for example for the $^{100}$Sn region.
\end{abstract}

\begin{keyword}
fusion-evaporation reactions \sep comparison experiment - calculations

%\PACS 
%25.70.Jj \sep 29.38.-c \sep 24.10.-i

\end{keyword}

\end{frontmatter}

\renewcommand{\footnotesep}{0.2mm}
\renewcommand{\footnoterule}{\rule{85mm}{.1mm}\vspace{2mm}}
\footnotetext[1]{e-mail address: blank@cenbg.in2p3.fr}  
\footnotetext[2]{Summer student at CENBG}


\section{Introduction}

On Earth, 255 stable nuclides are available for nuclear physics studies. In addition, 31 quasi stable nuclides having a half-life comparable
to or longer than the age of the Earth exist. All other nuclei must be created in order to be usable for experimental studies. Different types of nuclear reactions
exist to produce these unstable and radioactive nuclei. 

Two methods can be used to create basically all bound or quasi bound (i.e. bound for a 
short laps of time) nuclei: spallation or fragmentation. Spallation reactions are usually induced by light particles (protons or neutrons) 
on heavier stable nuclei. In these spallation reactions, the incident light projectile ejects nucleons from the target nucleus by nucleon-nucleon 
collisions and the excited fragment (often called pre-fragment) evaporates light particles (protons, neutrons, $\alpha$ particles) to get rid of
excitation energy. With e.g. incident proton energies of a few hundred MeV up to 1 or 2 GeV, basically all nuclei, bound or quasi bound, but lighter
than the target nucleus itself, can be produced. However, as these spallation reactions are basically always "thick-target" reactions, the reaction
products have to diffuse out of the target to become useful. As this takes some time and depends very sensitively on the chemistry of the element of
interest, short-lived nuclides of condensable elements are very difficult to produce by this means.

Fragmentation reactions employ heavy-ion induced reactions on different heavy-ion targets. Therefore, target as well as projectile fragmentation
can be used. Target fragmentation suffers from the same problem as spallation reactions: the products have to diffuse from the target itself.
Therefore, this process is again limited to relatively volatile isotopes with sufficiently long half-lives. In projectile fragmentation reactions, one can use "thin targets"
which allows the products to recoil out of the target due to the incident projectile energy. This approach is basically universal and allows all
nuclides to be produced. However, there are at least two drawbacks of projectile fragmentation: i) it needs high-energy heavy-ion accelerators and ii)
the beam quality of these fragment beams is rather bad.

In deep-inelastic or transfer reactions, two heavy nuclei interact with each other at energies around the Fermi energy (typically 20-60 MeV/A) 
and nucleons are transferred from one nucleus to the other producing thus more or less neutron-rich or neutron-deficient isotopes. However, 
as the number of nucleons transferred is limited, only nuclei relatively close to stability can be produced.

In nuclear fission, a very heavy nucleus, e.g. $^{238}$U or $^{252}$Cf, fissions by creating two medium-mass nuclides. This fission process can be
induced (e.g. by proton, neutron or $\gamma$-ray impact) or spontaneous. Due to the curvature of the nuclear
valley of stability, the heavy fissioning nuclei have always an excess of neutrons compared to lighter nuclei. Therefore, nuclear fission always
produces neutron-rich isotopes in the mass range of A~$\approx$ 50~- 170.

Finally, neutron-deficient nuclides can be produced by fusing two lighter nuclei. In this case, the situation is reversed compared to fission.
The light stable nuclei that interact are proton-rich compared to the heavier nuclei in the valley of stability. For example, the reaction of 
a stable $^{40}$Ca nucleus with a stable $^{58}$Ni nucleus produces as the compound nucleus, i.e. the sum of all nucleons, $^{98}$Cd, a nucleus
which is 8 neutrons more neutron-deficient than the most neutron-deficient stable isotope of the element cadmium.	

From this list of possible reactions, it is evident that the experimenter has some choice to use the reaction best suited for the production of the nucleus
of interest. However, evidently this choice depends also strongly on the accelerator available, the separation possibilities and much more. For each
type of reaction, parameters like the reaction partners and the incident energy have to be optimized in order to achieve the highest production rates
of the isotope of interest. For spallation, fragmentation, and deep-inelastic reactions, it is most often advantageous to use a stable nucleus close
to the desired final nucleus to enhance the production rate. This choice basically does not exist of fission because only a few quasi stable
fissioning nuclei exist. For these reactions, analytical codes have been developed which have a rather good predictive power for the reaction cross
sections. Let us mention the EPAX code~\cite{epax1,epax2,epax3} for projectile fragmentation, the SPACS code~\cite{spacs,spacs1} for spallation reactions or
the GRAZING model~\cite{grazing1,grazing2} for deep-inelastic reactions. The ABRABLA~\cite{gaimard91,kelic00} code deals with fission, fragmentation, and spallation.

For fusion-evaporation, the situation is different in the sense that all nuclei can be produced with different combinations of projectile, target and 
incident energy. Therefore, an optimization of these three parameters is needed for any nucleus to be produced. To do so, different codes are available, 
some of them being analytical, others being of the Monte-Carlo type. In the present work, we have used five codes to calculate fusion-evaporation cross sections:
CASCADE~\cite{cascade,cascade1}, HIVAP~\cite{hivap}, POTFUS+ABLA called CNABLA~\cite{cnabla,hagino99,gaimard91}, PACE~\cite{gavron80}, and 
POTFUS+GEMINI++~\cite{hagino99,gemini08}. All codes have advantages and draw-backs and we could not decide a priori which code would perform better 
over a wide range of nuclei.

The original reason for the present work was to determine production rates for SPIRAL2 where fusion-evaporation reactions were foreseen as a tool to
produce neutron-deficient isotopes from mass 20 or so to the heaviest nuclei in the super-heavy element region by means of a target - ion-source
ensemble in the production building. However, due to financial constraints, the construction of the SPIRAL2 production building was put on hold.
The same work was used in the mean time to predict production rates for the S$^3$ separator~\cite{s3} or at other facilities.

For this purpose, we performed a literature research of all fusion-evaporation reactions used to produce proton-rich nuclei. Using the projectile-target
combination and the energy given in the literature, cross sections were calculated with the five codes. To predict SPIRAL2 production rates, the in-target
yields were determined using the predicted primary-beam intensities and the extracted yields were obtained by means of release functions found in the literature.
In this way production rates could be predicted for more than 700 proton-rich nuclei.

In order to evaluate the performance of the fusion-evaporation codes and the quality of the production rate predictions, we have performed 
calculations with all five codes and compared the results to either fusion-evaporation cross sections found in the literature or to production 
rates from the GSI on-line separator~\cite{gsi-separator81}. The former values constitute a more direct comparison, however, in most cases 
the authors had to use transmissions of their separators which contain quite some uncertainties. For the second data, release efficiencies 
are needed in order to determine in-target production rates and thus production cross sections. To compare our cross-section calculations
with these values we will use release data collected in the frame work of the SPIRAL2 facility~\cite{spiral2} where these release functions are needed
for fusion as well as fission products.

The purpose of the present paper is to describe the results of this comparison between calculated cross sections or production rates and
experimental data for fusion-evaporation reactions. The general outcome is that the different codes overestimate the experimental data by about a factor of 10. Therefore, 
for e.g. planning an experiment using a fusion-evaporation reaction, the predictions deduced from calculations using fusion-evaporation codes
should be reduced by this factor in order to obtain a realistic estimate of the production rates to be expected.

\section{Experimental data}

In this section, we summarize the experimental data used for the comparison with the theoretical predictions. Table~\ref{tab:exp-cross} gives
the experimental cross sections used in the present work~\footnote[3]{The authors are eager to increase the present data base of experimental cross sections
and encourage readers to communicate other experimental fusion-evaporation cross sections to us.}.

In general, relatively few fusion-evaporation cross sections are found in the literature and those found have often large error bars or, even worse,
no uncertainties at all. This is to a large part due to the fact that the cross sections are often determined at ISOL facilities where 
information of effusion and diffusion is scarce and induce large uncertainties. Other cross sections are determined by means of mass 
separators or velocity filters where the transmissions are not well known.

\begin{table}
\caption{Experimental cross sections from literature. Given are the mass and charge number of the nuclei of interest, the mass and charge number of the
         projectile and target nuclei, respectively, the incident beam energy, the experimental cross section and its error if available, and the reference. }
{\renewcommand{\arraystretch}{0.96}%  1 is the default, change whatever you need
\vspace*{3mm}
\begin{tabular}{r@{\hspace{2.8mm}}r@{\hspace{2.8mm}}r@{\hspace{2.8mm}}r@{\hspace{2.8mm}}r@{\hspace{2.8mm}}r@{\hspace{2.8mm}}r@{\hspace{2.8mm}}c@{\hspace{4mm}}c@{\hspace{4mm}}l}
 A & Z & A$_p$ & Z$_p$ & A$_t$ & Z$_t$ & \multicolumn{1}{c}{E} & cross & error & Ref.  \\
   &   &       &       &       &       &\multicolumn{1}{c}{(MeV)}& section & (mb) &       \\
   &   &       &       &       &       &   & (mb) &    &       \\
\hline
\multicolumn{10}{l}{light N$\approx$Z nuclei:} \\
64	&  30  & 12 &  6  &   54 &  26 &      37  &      1.60E+02  &    7.00E+00                  &    \cite{ennis91}     \\
64	&  31  & 54 &  26 &   12 &  6  &     150  &      7.90E+01  &                              &    \cite{ooi86}     \\
64	&  32  & 40 &  20 &   27 &  13 &     102  &      4.00E-01  &    6.00E-02                  &    \cite{goerres87}      \\      
64	&  32  & 54 &  26 &   12 &  6  &     165  &      6.40E-01  &    7.00E-02                  &    \cite{ennis91}     \\      
64	&  32  & 54 &  26 &   12 &  6  &     150  &      3.40E-01  &    9.00E-02                  &    \cite{ooi86}     \\
64	&  32  & 54 &  26 &   12 &  6  &     165  &      5.00E-01  &    3.00E-01                  &    \cite{lister90}    \\    
64	&  32  & 12 &  6  &   58 &  28 &     40   &      2.00E-01  &    5.00E-02                  &    \cite{skoda98}       \\
68	&  34  & 58 &  28 &   12 &  6  &     175  &      3.80E-02  &    1.60E-02                  &    \cite{lister90}    \\
68	&  34  & 58 &  28 &   12 &  6  &     220  &      2.00E-01  &    5.00E-02                  &    \cite{fischer00}     \\
72	&  36  & 16 &  8  &   58 &  28 &     55   &      1.00E-01  &    3.00E-02                  &    \cite{skoda98}       \\
72	&  36  & 58 &  28 &   16 &  8  &     170  &      6.00E-02  &    2.50E-02                  &    \cite{varley87}     \\
76	&  38  & 54 &  26 &   24 &  12 &     175  &      1.00E-02  &    5.00E-03                  &    \cite{lister90}    \\
80	&  40  & 58 &  28 &   24 &  12 &     190  &      1.00E-02  &    5.00E-03                  &    \cite{lister87}     \\
80	&  39  & 58 &  28 &   24 &  12 &     190  &      2.00E+00  &    1.00E+00                  &    \cite{lister87}     \\
80	&  38  & 58 &  28 &   24 &  12 &     190  &      4.40E+01  &    4.00E+00                  &    \cite{lister87}     \\
\multicolumn{10}{l}{$^{100}$Sn region:} \\                                                       
 95	&  45  & 58 &  28 &   50 &  24 &     250  &      1.10E+00  &    4.00E-01                  &    \cite{schubart95}     \\
 97	&  45  & 58 &  28 &   50 &  24 &     250  &      3.40E+00  &    2.00E-01                  &    \cite{schubart95}     \\
 98	&  46  & 58 &  28 &   50 &  24 &     250  &      2.20E+01  &    2.00E+00                  &    \cite{schubart95}     \\
 98	&  47  & 58 &  28 &   50 &  24 &     250  &      3.00E-01  &    6.00E-02                  &    \cite{schubart95}     \\
 99	&  47  & 58 &  28 &   50 &  24 &     250  &      3.60E+00  &    4.00E-01                  &    \cite{schubart95}     \\
 99 &  48  & 58 &  28 &   50 &  24 &     249  &      3.20E-02  &    2.00E-02                  &    \cite{lacommara00}      \\
 99 &  48  & 58 &  28 &   50 &  24 &     249  &      3.20E-02  &    2.00E-02                  &    \cite{lacommara00}      \\
 99 &  48  & 50 &  24 &   58 &  28 &     225  &      2.50E-02  &    8.00E-03                  &    \cite{lacommara00}      \\
 99 &  48  & 58 &  28 &   58 &  28 &     348  &      1.10E-02  &    8.00E-03                  &    \cite{lacommara00}      \\
 99 &  48  & 58 &  28 &   58 &  28 &     371  &      2.80E-02  &    2.10E-02                  &    \cite{lacommara00}      \\
 99 &  48  & 58 &  28 &   58 &  28 &     394  &      3.10E-02  &    2.00E-02                  &    \cite{lacommara00}      \\
100	&  47  & 58 &  28 &   50 &  24 &     250  &      3.90E+00  &    2.00E-01                  &    \cite{schubart95}    \\
100	&  47  & 50 &  24 &   58 &  28 &     225  &      3.90E+00  &                              &    \cite{schubart95}    \\
100	&  48  & 50 &  24 &   58 &  28 &     225  &      1.00E+00  &                              &    \cite{chartier96}     \\
100	&  49  & 50 &  24 &   58 &  28 &     225  &      1.00E-03  &                              &    \cite{chartier96}     \\
100 &  49  & 58 &  28 &   50 &  24 &     319  &      2.60E-03  &                              &    \cite{lacommara00}      \\
100 &  49  & 58 &  28 &   58 &  28 &     325  &      8.00E-04  &                              &    \cite{lacommara00}      \\
100 &  49  & 58 &  28 &   58 &  28 &     348  &      1.70E-03  &                              &    \cite{lacommara00}      \\
100 &  49  & 58 &  28 &   58 &  28 &     371  &      1.70E-03  &                              &    \cite{lacommara00}      \\
100 &  49  & 58 &  28 &   58 &  28 &     394  &      1.60E-03  &                              &    \cite{lacommara00}      \\
100	&  50  & 50 &  24 &   58 &  28 &     225  &      4.00E-05  &                              &    \cite{chartier96}     \\
101	&  47  & 58 &  28 &   50 &  24 &     250  &      4.70E+01  &    3.00E+00                  &    \cite{schubart95}     \\
101	&  48  & 58 &  28 &   50 &  24 &     250  &      1.80E+01  &    2.00E+00                  &    \cite{schubart95}     \\
101 &  50  & 58 &  28 &   50 &  24 &     249  &      1.60E-05  &    4.00E-06                  &    \cite{lacommara00}      \\
101 &  50  & 58 &  28 &   50 &  24 &     250  &      1.00E-05  &                              &    \cite{lacommara00}      \\
101 &  50  & 58 &  28 &   58 &  28 &     325  &      9.00E-06  &    4.00E-06                  &    \cite{lacommara00}      \\
101 &  50  & 58 &  28 &   58 &  28 &     348  &      1.30E-05  &    3.00E-06                  &    \cite{lacommara00}      \\
101 &  50  & 58 &  28 &   58 &  28 &     371  &      2.80E-05  &    1.00E-05                  &    \cite{lacommara00}      \\
101 &  50  & 58 &  28 &   58 &  28 &     394  &      7.00E-06  &    4.00E-06                  &    \cite{lacommara00}      \\
102	&  48  & 58 &  28 &   50 &  24 &     250  &      6.30E+01  &    1.90E+01                  &    \cite{schubart95}     \\
102 &  49  & 58 &  28 &   50 &  24 &     249  &      9.00E-01  &    5.00E-01                  &    \cite{lacommara00}      \\
102 &  49  & 58 &  28 &   50 &  24 &     249  &      1.30E+00  &    7.00E-01                  &    \cite{lacommara00}      \\
102 &  49  & 58 &  28 &   50 &  24 &     348  &      1.10E+00  &    6.00E-01                  &    \cite{lacommara00}      \\
102 &  49  & 58 &  28 &   58 &  28 &     325  &      1.20E+00  &    6.00E-01                  &    \cite{lacommara00}      \\
102 &  49  & 58 &  28 &   58 &  28 &     348  &      1.20E+00  &    6.00E-01                  &    \cite{lacommara00}      \\
102 &  49  & 58 &  28 &   58 &  28 &     348  &      7.00E-01  &    3.00E-01                  &    \cite{lacommara00}      \\
102 &  49  & 58 &  28 &   58 &  28 &     371  &      1.00E+00  &    5.00E-01                  &    \cite{lacommara00}      \\
102 &  49  & 58 &  28 &   58 &  28 &     394  &      9.00E-01  &    4.00E-01                  &    \cite{lacommara00}      \\
102	&  50  & 58 &  28 &   52 &  24 &     225  &      2.00E-03  &                              &    \cite{lipoglavsek96}     \\
103	&  47  & 58 &  28 &   50 &  24 &     250  &      3.60E+00  &    4.00E-01                  &    \cite{schubart95}     \\
103	&  48  & 58 &  28 &   50 &  24 &     250  &      2.70E+01  &    2.00E+00                  &    \cite{schubart95}     \\
103	&  49  & 58 &  28 &   50 &  24 &     250  &      6.40E+00  &    8.00E-01                  &    \cite{schubart95}     \\
104	&  48  & 58 &  28 &   50 &  24 &     250  &      1.79E+02  &    7.00E+00                  &    \cite{schubart95}     \\
104	&  49  & 58 &  28 &   50 &  24 &     250  &      5.80E+01  &    1.60E+01                  &    \cite{schubart95}     \\
104	&  50  & 58 &  28 &   50 &  24 &     250  &      1.80E+00  &    2.00E-01                  &    \cite{schubart95}     \\
\end{tabular}
}
\label{tab:exp-cross}
\end{table}
\begin{table}
{\renewcommand{\arraystretch}{0.96}%  1 is the default, change whatever you need
\begin{tabular}{r@{\hspace{2.8mm}}r@{\hspace{2.8mm}}r@{\hspace{2.8mm}}r@{\hspace{2.8mm}}r@{\hspace{2.8mm}}r@{\hspace{2.8mm}}r@{\hspace{2.8mm}}c@{\hspace{4mm}}c@{\hspace{4mm}}l}
 A & Z & A$_p$ & Z$_p$ & A$_t$ & Z$_t$ & \multicolumn{1}{c}{E} & cross & error & Ref.  \\
   &   &       &       &       &       &\multicolumn{1}{c}{(MeV)}& section & (mb) &       \\
   &   &       &       &       &       &   & (mb) &    &       \\
\hline
105	&  49  & 58 &  28 &   50 &  24 &     250  &      1.16E+02  &    6.00E+00                  &    \cite{schubart95}     \\
105	&  50  & 58 &  28 &   50 &  24 &     250  &      1.00E+01  &    2.00E+00                  &    \cite{schubart95}     \\
\multicolumn{10}{l}{Ba nuclei:} \\                                                               
114 &  56  & 58 &  28 &    58&  28 &   222-248&      2.00E-04  &    1.00E-04                  &    \cite{mazzocchi01}      \\
114 &  56  & 58 &  28 &    58&  28 &   203-244&      2.00E-04  &    1.00E-04                  &    \cite{janas97}     \\
116 &  56  & 58 &  28 &    60&  28 &   209-249&      3.00E-03  &    1.00E-03                  &    \cite{janas97}     \\
116 &  56  & 58 &  28 &    63&  29 &   249-284&      8.00E-04  &    4.00E-04                  &    \cite{janas97}     \\
117 &  56  & 58 &  28 &    63&  29 &   249-284&      5.50E-02  &    2.00E-02                  &    \cite{janas97}     \\
118 &  56  & 58 &  28 &    63&  29 &   249-284&      1.90E-02  &    6.00E-03                  &    \cite{janas97}     \\
\multicolumn{10}{l}{heavier nuclei:} \\                  
171 &  79  & 78 &  36 &    96&  44 &      361 &      1.10E-03  &                             &     \cite{kettunen04}   \\
171 &  79  & 78 &  36 &    96&  44 &      359 &      2.00E-03  &                             &     \cite{kettunen04}   \\
171 &  79  & 78 &  36 &    96&  44 &      363 &      6.00E-04  &                             &     \cite{kettunen04}   \\
170 &  79  & 78 &  36 &    96&  44 &      386 &      9.00E-05  &                             &     \cite{kettunen04}   \\
173 &  80  & 78 &  36 &   102&  46 &      384 &      4.00E-06  &                             &     \cite{kettunen04}   \\
172 &  80  & 78 &  36 &    96&  44 &      361 &      4.00E-06  &                             &     \cite{kettunen04}   \\
171 &  80  & 78 &  36 &    96&  44 &      361 &      2.00E-06  &                             &     \cite{kettunen04}   \\
176 &  81  & 78 &  36 &   102&  46 &      384 &      3.00E-06  &                             &     \cite{kettunen04}   \\
172 &  80  & 78 &  36 &    96&  44 &      375 &      9.00E-06  &                             &     \cite{seweryniak99}   \\
173 &  80  & 80 &  36 &    96&  44 &      400 &      1.50E-05  &                             &     \cite{seweryniak99}   \\
174 &  80  & 80 &  36 &    96&  44 &      375 &      3.30E-04  &                             &     \cite{seweryniak99}   \\
\multicolumn{10}{l}{proton emitter: pn channel:} \\                      
185  & 83  &  92  &  42  &  95   & 42  &   410    &  1.00E-04  &                             &   \cite{davids96}           \\            
185  & 83  &  92  &  42  &  95   & 42  &   420    &  6.00E-05  &                             &   \cite{poli01}       \\                 
\multicolumn{10}{l}{p2n channel:} \\                     
109	 & 53  &  58  &  28  &  54   & 26  &   195    &  1.00E-02  &                             &   \cite{petri07}        \\               
109	 & 53  &  58  &  28  &  54   & 26  &   220    &  1.60E-02  & 4.00E-03                    &   \cite{yu99}         \\                 
109	 & 53  &  58  &  28  &  54   & 26  &   240    &  3.00E-03  &                             &   \cite{paul95}            \\            
109	 & 53  &  58  &  28  &  54   & 26  &   229    &  5.00E-02  &                             &   \cite{hofmann89} \\                    
109	 & 53  &  58  &  28  &  54   & 26  &   250    &  4.00E+01  & $^{+4.00E+01}_{-2.00E+01}$  &   \cite{faestermann84}          \\       
109	 & 53  &  58  &  28  &  58   & 28  &   250    &  3.00E+01  & $^{+3.00E+01}_{-1.50E+01}$  &   \cite{faestermann84}          \\       
113	 & 55  &  58  &  28  &  58   & 28  &   250    &  3.00E+01  &                             &   \cite{faestermann84}         \\        
147  & 69  &  58  &  28  &  92   & 42  &   260    &  1.80E-02  &                             &   \cite{seweryniak97}         \\         
151	 & 71  &  58  &  28  &  96   & 44  &   266    &  7.00E-02  & 1.00E-02                    &   \cite{yu98}                       \\              
161	 & 75  &  58  &  28  &  106  & 48  &   270    &  6.30E-03  & 1.80E-03                    &   \cite{irvine97}         \\             
167	 & 77  &  78  &  36  &  92   & 42  &   357    &  1.10E-01  &                             &   \cite{davids97}          \\            
171	 & 79  &  78  &  36  &  96   & 44  &   389    &  2.00E-03  &                             &   \cite{davids97}          \\            
171  & 79  &  78  &  36  &  96   & 44  &   370    &  6.00E-04  &                             &   \cite{baeck03}          \\             
171  & 79  &  78  &  36  &  96   & 44  &   361    &  1.10E-03  &                             &   \cite{kettunen04}        \\            
171  & 79  &  78  &  36  &  96   & 44  &   359    &  2.00E-03  &                             &   \cite{kettunen04}        \\            
171  & 79  &  78  &  36  &  96   & 44  &   363    &  6.00E-04  &                             &   \cite{kettunen04}        \\            
177	 & 81  &  78  &  36  &  102  & 46  &   370    &  3.00E-05  &                             &   \cite{poli99}        \\                
\multicolumn{10}{l}{p3n channel:} \\                            
108  & 53  &  58  &  28  &  54   & 26  &  240-255 &  5.00E-04  &                             &   \cite{page94a}          \\              
112  & 55  &  58  &  28  &  58   & 28  &   259    &  5.00E-04  &                             &   \cite{page94}         \\              
146	 & 69  &  58  &  28  &  92   & 42  &   287    &  1.00E-03  &                             &   \cite{livingston93}           \\       
150	 & 71  &  58  &  28  &  96   & 44  &   297    &  2.56E-03  &                             &   \cite{robinson03}        \\            
150	 & 71  &  58  &  28  &  96   & 44  &   292    &  3.05E-03  &                             &   \cite{ginter99}        \\              
160	 & 75  &  58  &  28  & 106   & 48  &   300    &  1.00E-03  &                             &   \cite{page92}          \\              
166	 & 77  &  78  &  36  &  92   & 42  &   384    &  6.30E-03  &                             &   \cite{davids97}          \\            
176  & 81  &  78  &  36  & 102   & 46  &   384    &  3.00E-06  &                             &   \cite{kettunen04}        \\            
\multicolumn{10}{l}{p4n channel:}  \\                           
117	 & 57  &  58  &  28  &  64   & 30  &   310    &  2.00E-04  &                             &   \cite{soramel01}       \\              
117	 & 57  &  58  &  28  &  64   & 30  &  295,310 &  2.40E-04  & $^{+2.40E-04}_{-1.20E-04}$  &   \cite{mahmud01}       \\               
131	 & 63  &  40  &  20  &  96   & 44  &   222    &  9.00E-05  &                             &   \cite{davids98}          \\            
141	 & 67  &  54  &  26  &  92   & 42  &  285,305 &  2.50E-04  &                             &   \cite{davids98}           \\           
141	 & 67  &  54  &  26  &  92   & 42  &   315    &  3.00E-05  &                             &   \cite{rykaczewski99}        \\         
145  & 69  &  58  &  28  &  92   & 42  &   315    &  5.00E-04  &                             &   \cite{batchelder98}         \\         
145  & 69  &  92  &  42  &  58   & 28  &   512    &  2.00E-04  &                             &   \cite{seweryniak05}          \\        
155	 & 73  &  58  &  28  & 102   & 46  & 315,320  &  6.00E-05  &                             &   \cite{uusitalo99}         \\           
165	 & 77  &  78  &  36  &  92   & 42  &   384    &  2.00E-04  &                             &   \cite{davids97}          \\            
\multicolumn{10}{l}{p5n channel:} \\                            
130  & 63  &  78  &  36  &  58   & 28  &   432    &  9.00E-06  &  $^{+9.00E-06}_{-4.50E-06}$ &   \cite{davids04}     \\                 
140	 & 67  &  54  &  26  &  92   & 42  &   315    &  3.00E-06  &                             &   \cite{rykaczewski99}        \\         
\multicolumn{10}{l}{p6n channel:} \\                            
121  & 59  &  36  &  18  &  92   & 42  &   240    &  3.00E-07  &  $^{+3.00E-07}_{-1.(0E-07}$ &   \cite{robinson05}        \\            
135  & 65  &  50  &  24  &  92   & 42  &   310    &  3.00E-06  &                             &   \cite{woods04}        \\               
\end{tabular}
}
\end{table}

Another problem with a comparison of experimental cross sections and calculated values is that it is often not clear whether the beam energy given 
is the one at the entrance or in the center of the target. We always use the energy given in the paper for the calculations. 
If the energy is the one at the target entrance and thus too high compared to the energy in the center of the target, we believe this in not a problem. 
The maximum of the cross sections is reached at a certain incident energy. At higher energies, the cross sections fall off slowly, whereas at lower 
energies there is a threshold effect to overcome the Coulomb repulsion which makes that the cross sections fall off much faster on the low-energy side. 
Therefore, taking in some cases a slightly higher beam energy is somehow on the "safe" side.

Experimental production rates can be found in a number of publications from the former GSI on-line separator~\cite{kirchner81}.
They are summarized in table~\ref{tab:exp-prod}.

\begin{table}
\caption{Experimental production rates from literature. Given are the mass and charge number of the nuclei of interest and the reference. }
{\renewcommand{\arraystretch}{0.96}%  1 is the default, change whatever you need
\vspace*{3mm}
\begin{center}
\begin{tabular}{c@{\hspace{5mm}}c@{\hspace{5mm}}l@{\hspace{20mm}}c@{\hspace{5mm}}c@{\hspace{5mm}}l@{\hspace{5mm}}}
 A  & Z   &  Ref.  &  A  & Z   &  Ref.  \\
\hline
 60 & 31  & \cite{mazzocchi01}              & 105 & 49  & \cite{karny01}              \\
 61 & 31  & \cite{oinonen99}                & 106 & 49  & \cite{palit10,deo09}        \\
 62 & 31  & \cite{blank04ga62}              & 107 & 49  & \cite{karny01}              \\
 94 & 47  & \cite{plettner04,lacommara02}   & 101 & 50  & \cite{lacommara00}          \\
 95 & 47  & \cite{harissopulos05}           & 102 & 50  & \cite{lipoglavsek96}        \\
 96 & 47  & \cite{mukha04}                  & 103 & 50  & \cite{kavatsyuk05}          \\
 97 & 47  & \cite{hu99}                     & 104 & 50  & \cite{schubart95}           \\
 98 & 47  & \cite{hu00}                     & 105 & 50  & \cite{schubart95}           \\
100 & 49  & \cite{lacommara00}              & 114 & 56  & \cite{janas97,mazzocchi02}  \\
102 & 49  & \cite{lacommara00,gierlik03}    & 116 & 56  & \cite{janas97}              \\
103 & 49  & \cite{szerypo97}                & 117 & 56  & \cite{janas97}              \\
104 & 49  & \cite{karny01}                  & 118 & 56  & \cite{janas97}              \\
\end{tabular}
\end{center}
}
\label{tab:exp-prod}
\end{table}

\section{Simulation codes}

In this section, we give a short overview of the fusion-evaporation codes used to calculate the theoretical cross sections. In total, 
five codes were used: i) CASCADE, ii) HIVAP, iii) CNABLA, iv) PACE, and v) GEMINI++. 
These codes use a two-step scenario for the reaction: projectile and target nuclei completely fuse and then decay according to a statistical model approach
of compound nucleus reactions. They take into account competition between different decay channels like proton, neutron, and $\alpha$ emission as well as 
$\gamma$ decay and fission. All codes give a variety of decay information like the particles emitted, their energy and angular distribution etc.
In the present work, we only use the production cross section of the isotope of interest. All programs used the Atomic Mass Evaluation data base from 
2012~\cite{ame2012}.

\subsection{The CASCADE code}

The program CASCADE was originally written by F. P{\"u}hlhofer~\cite{cascade}.
The original version of the program was modified by different persons (e.g. E.F. Garman, F. Zwarts and M.N. Harakeh) to perform calculations for special 
states of good spin and parity, to include isospin and parity properly in the statistical decay as well as to include the electric quadrupole decay.

CASCADE is an analytic program which is quite fast and thus convenient to optimize projectile-target combinations and the beam energy.
In the present work, we use a version of CASCADE provided by D. R. Chakrabarty~\cite{cascade1}.

\subsection{The HIVAP code}

HIVAP is a statistical evaporation code written by W. Reisdorf~\cite{hivap}. Several improvements were introduced later~\cite{reisdorf85,reisdorf92}.
We used a version provided to us by F. Hessberger~\cite{hessberger00}. Like CASCADE, HIVAP is an analytical program being thus very fast.

\subsection{The CNABLA code}

CNABLA is a program which combines the POTFUS fusion code~\cite{hagino99} for the first step of the reaction with the ABLA part from the ABRABLA code~\cite{gaimard91}
for the evaporation. POTFUS is a quite successfully used fusion code and allows us to prepare an input file with a predefined number of events
with four parameters: the mass and the charge of the complete-fusion product, its excitation energy and its spin. These events are then used with
a special version of ABRABLA~\cite{kelic00} to perform the evaporation part by means of a Monte-Carlo technique.

\subsection{The PACE code}

PACE is probably the most widely used fusion-evaporation code. It was originally written by A. Gavron~\cite{gavron80}. This Projection Angular-momentum Coupled Evaporation (PACE)
code is again based on the statistical model and uses the Monte-Carlo approach for the de-excitation of the compound nucleus. Only the equilibrium part of the decay
is treated, no pre-equilibrium emission is considered.

\subsection{The GEMINI++ code}

The GEMINI++ code~\cite{gemini08} is the C++ version of the original GEMINI code~\cite{charity88,charity90} written by R.J. Charity. In addition to light particle emission and symmetric fission, 
it allows for all binary decays to occur. This new version cures problems with heavier systems in the original code. The complete fusion compound nuclei are again produced by the POTFUS 
code~\cite{hagino99} and read into GEMINI++ where a Monte-Carlo procedure is used to perform the de-excitation step.

\begin{figure*}[hht]
\vspace*{-2cm}
\begin{center}
\includegraphics[width=0.95\textwidth,angle=0]{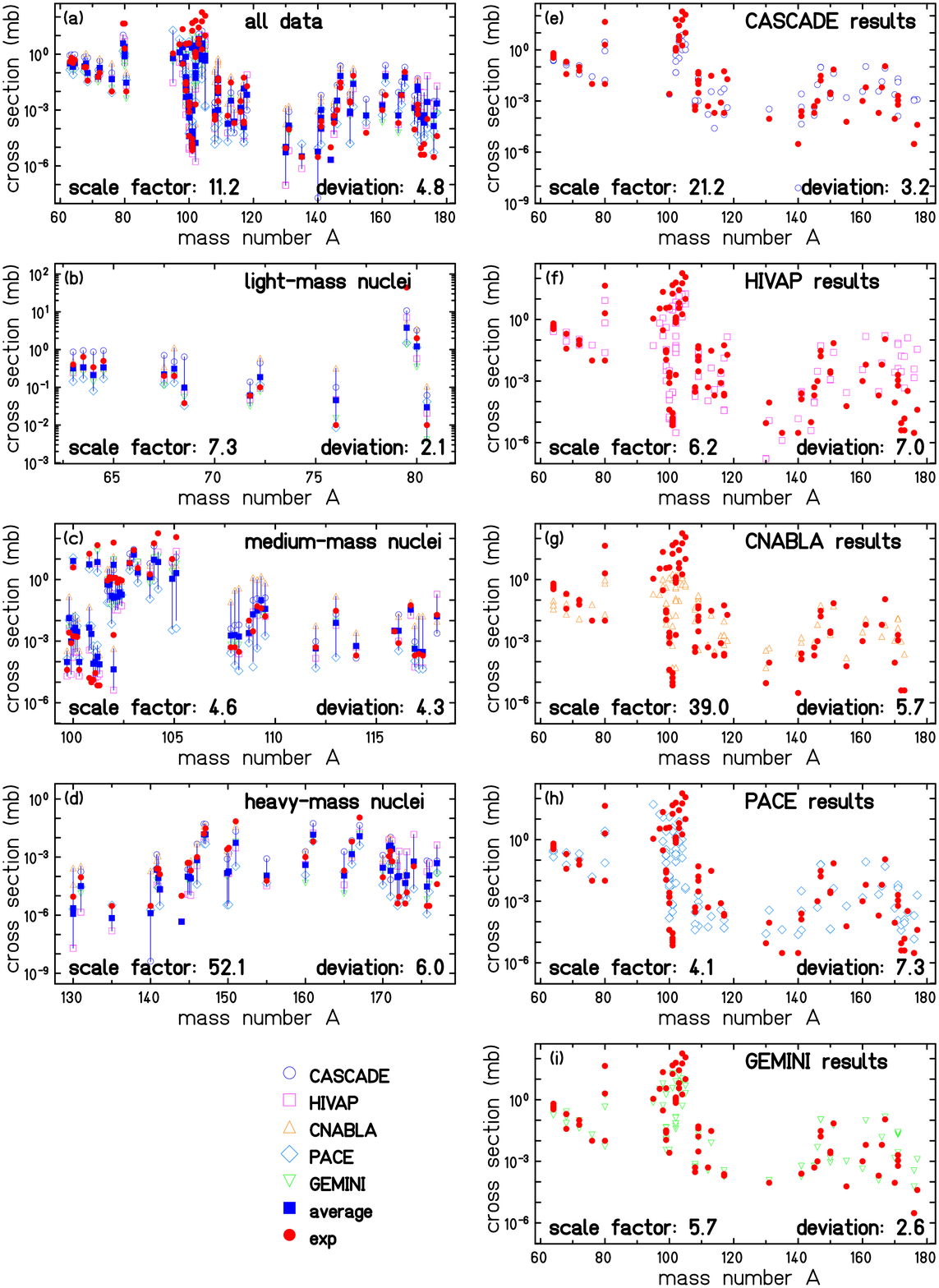}
\caption{Comparison of experimental cross sections taken from Table~\ref{tab:exp-cross} and calculated cross sections with the five different models.
          Each figure gives the scale factor by which the calculated cross sections had to be divided to match the experimental cross sections.
          The deviation gives the average difference factor between the experimental cross sections and the scaled calculated cross sections (see text).
          The left column (a-d) compares the experimental cross sections to all fives model calculations, for all data (a) and for different mass ranges (b-d).
          The right column compares all experimental data with the different models (e-i).
         } 
\label{fig:cross_all}
\end{center}
\end{figure*}

\begin{figure*}[hht]
\begin{center}
\includegraphics[width=0.75\textwidth,angle=-90]{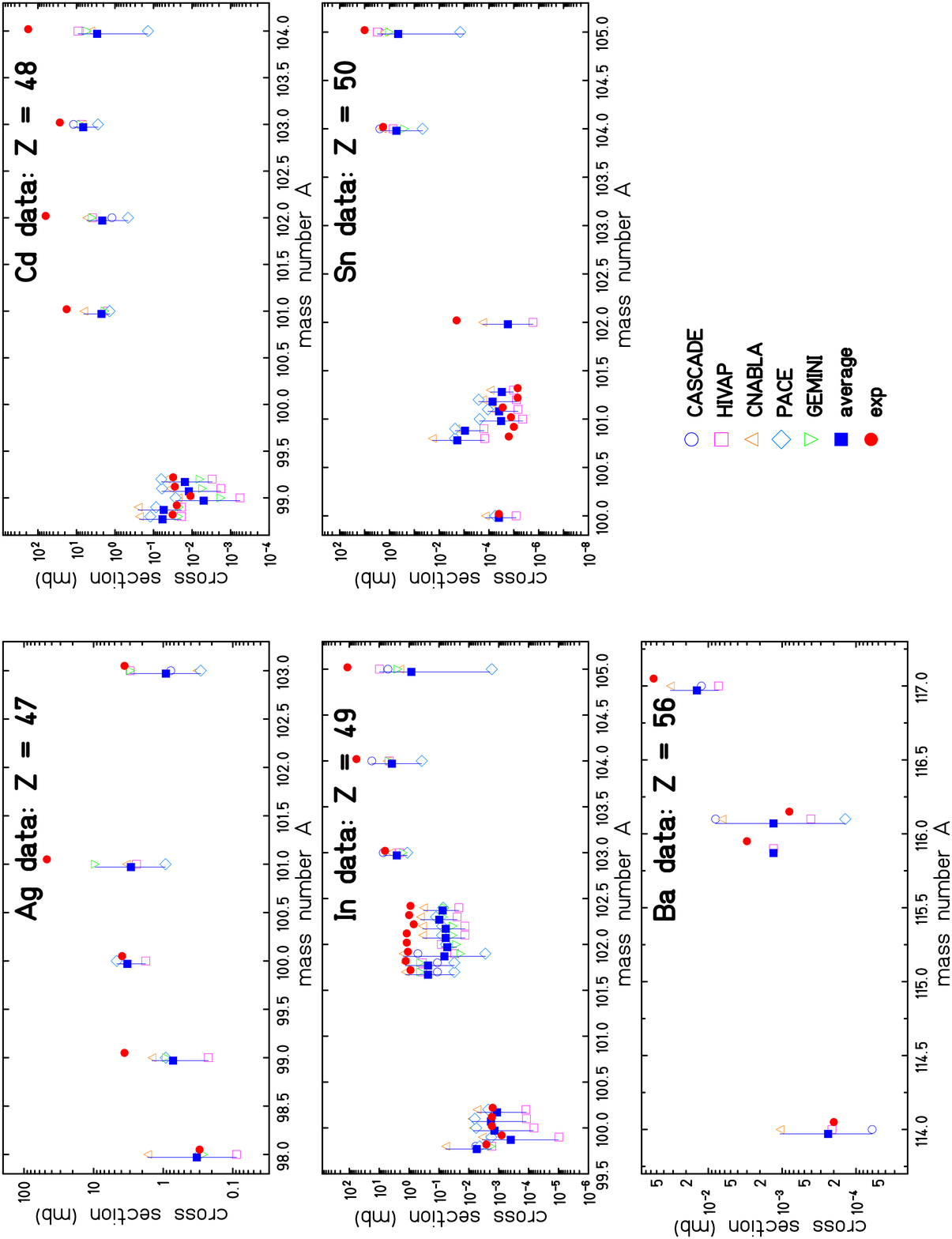}
\caption{Comparison of experimental and calculated cross sections (with the adopted scaling factor of 11.2) for selected elements.
         } 
\label{fig:cross_elem}
\end{center}
\end{figure*}

\subsection{Averages from calculations}

In order to compare the experimental results to the theoretical predictions from the five codes, some averaging of the calculations is needed.
This task is not so easy because the calculations can differ by one or two orders of magnitude from one code to another.
A standard average would favor the larger cross sections (e.g. the average of 1~mb and 100~mb being about 50~mb). Therefore, we decided to use the 
geometrical average yielding for the example above an average of 10~mb. As the uncertainty range we used the maximum and minimum value from all codes.

In general, not all codes give results for all isotopes or projectile, target, and energy combinations. The average is therefore made
with the results available.

\section{Results and discussion}
\subsection{Comparison with experimental cross sections}

Figure~\ref{fig:cross_all} gives an overview of all experimental data compared with the results of the individual codes and the averages of these
calculations as explained above. As indicated on the figure~\ref{fig:cross_all}a to get the best match between the average of the simulations and the experimental data, we had to 
reduce the results of the calculations by a scale factor of 11.2. The parameter called deviation is a measure for the scatter of the calculated cross sections, 
after scaling,  around the experimental ones. Again due to large differences between the calculated values from different models, we used a logarithmic difference 
defined as:
$$
deviation = 10^{**}\left[1/n \sum_n abs\left(log_{10} \left(\frac{\sigma_{cal}/sf}{\sigma_{exp}} \right)\right)\right]
$$
where $n$ is the number of data points, $\sigma_{cal}$ and $\sigma_{exp}$ are the calculated and the experimental cross sections, respectively,
and $sf$ is the scale factor mentioned above. Therefore, this deviation is the average factor by which the calculations deviate from the experimental value:
the smaller this value is, the better the model calculation, once scaled by a constant factor, agrees with experimental data.

From figure~\ref{fig:cross_all}a, we conclude that the average of the five model calculations corrected by a scale factor of 11.2 deviate on  average by a
factor of close to 5 for individual value. As can be seen from the left-hand side of the figure~\ref{fig:cross_all}, the scatter between the models and the experimental
data is much better for lighter nuclei and gets worse when moving to heavier nuclei.

\begin{figure*}[hht]
\begin{center}
\includegraphics[width=0.5\textwidth,angle=-90]{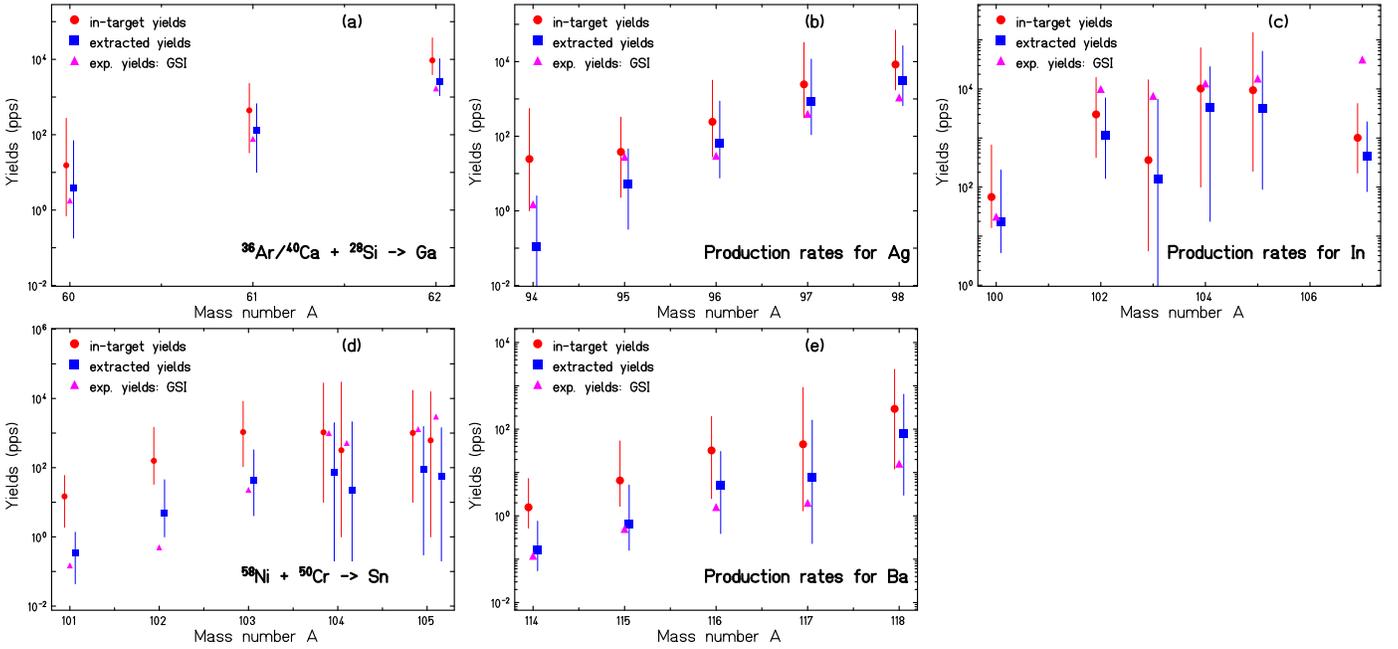}
\caption{Comparison of experimental and calculated production rates with a scale factor of 7.3 for the mass A=60 region (a) and of 4.6
         for the mass A=100 region (b-e).
         } 
\label{fig:prod}
\end{center}
\end{figure*}

The right-hand side of figure~\ref{fig:cross_all} gives an analysis of the results as a function of the model used to calculate the cross sections.
From a first glance, it seems that PACE is the best model, because the scale factor is the smallest of all. However, the scatter of the data is the largest
of all models. Overall we believe that the GEMINI++ model coupled to the POTFUS fusion program gives the most convincing answer for fusion-evaporation
cross sections. As in the other cases, the agreement is better for the low-mass region (A $<$ 90) with a scale factor of 3.4 and a deviation parameter
of 1.9 and for the medium mass region (90 $<$ A $<$ 130) with values of 2.8 and 1.9.

Interestingly the models which need a large scale factor to match the experimental data, CASCADE and CNABLA, have a reasonably small scatter of the data.
This is in particular true for the CASCADE model. HIVAP has a reasonably small scale factor but a very large scatter of the data.

\begin{table}[bb]
\caption{Scale factors for calculations and deviations between calculated values and experimental data for the five different models in the $^{100}$Sn region.}
{\renewcommand{\arraystretch}{0.96}%  1 is the default, change whatever you need
\vspace*{1mm}
\begin{center}
\begin{tabular}{c@{\hspace{8.0mm}}c@{\hspace{8.0mm}}c@{\hspace{8.0mm}}}
 model & scale factor & deviation  \\
\hline
CASCADE & ~6.3  &  1.9 \\
HIVAP   & ~1.9  &  3.5 \\
CNABLA  & 22.8  &  7.0 \\
PACE    & ~2.1  &  8.9 \\
GEMINI++& ~2.4  &  1.9 \\
\hline
\end{tabular}
\end{center}
}
\label{tab:sn100}
\end{table}

In the $^{100}$Sn region, a lot of experiments have been performed and experimental cross sections determined, notably at  the former GSI on-line 
separator~\cite{kirchner81}. Therefore, this region allows for a detailed comparison of experimental data and calculations.
If we use the overall scale factor of 11.2, we obtain a rather good match between experimental data and calculations for the most exotic nuclei
(see figure~\ref{fig:cross_elem}). However, closer to stability the experimental data are underestimated by the thus scaled calculations. This statement is
valid for all elements from silver (Z=47) to barium (Z=56).

An interesting question is certainly, which model predicts best cross sections in the $^{100}$Sn region. If we compare the large body of experimental data
from A=94 to A=117 to the different models, we get scale factors and deviations as given in Table~\ref{tab:sn100}. In this region, HIVAP and POTFUS+GEMINI++
give the best results with small scale factors and small deviations.

\subsection{Comparison with on-line production rates}

Another possibility to compare predictions and experimental rates is to use production rates 
achieved in experiments and compare them to calculated rates. This comparison is possible 
with production rates published from the former GSI on-line separator (see Table~\ref{tab:exp-prod}).
However, in such a comparison the uncertainties are expected to be even larger because, in 
order to calculate these rates, one has to make assumptions about release and ionization 
efficiencies. This is a rather difficult task, because it involves
a lot of chemistry and the on-line rates are known to fluctuate from one run to the other due 
to often apparently minor differences of the experimental conditions of different 
experiments.

\begin{figure*}[hht]
\begin{center}
\includegraphics[width=0.5\textwidth,angle=-90]{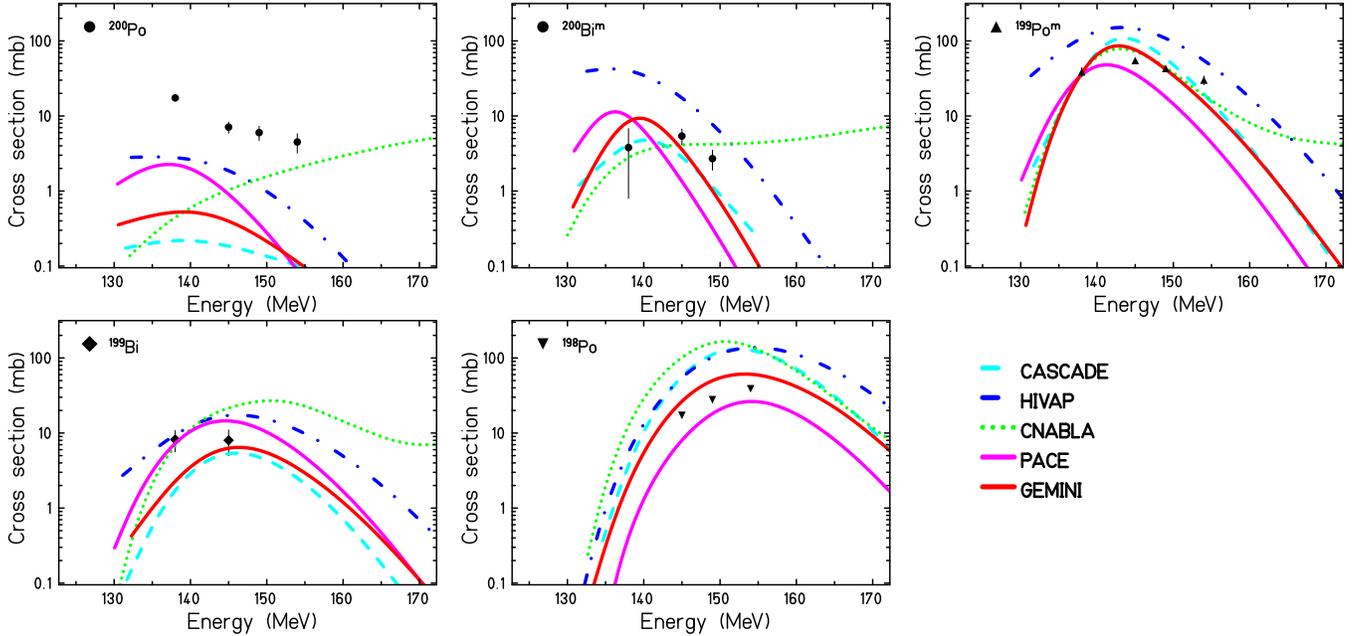}
\caption{Comparison of experimental excitation functions for nuclei in the mass A=200 region ~\cite{sudarshan17} with predictions from the five models used.
         The energy range is unfortunately too small to draw conclusions about the accordance of the maximum of the distributions between
         experimental data and models.
         } 
\label{fig:energy}
\end{center}
\end{figure*}

Nevertheless, we have attempted to predict production rates for the future SPIRAL2 facility 
at GANIL, be it for neutron-induced fission of $^{238}$U or fusion-evaporation reactions 
for proton-rich nuclei~\cite{spiral2webpage}. For this purpose, we have collected 
experimental parameters of two types: (i) empirical parameterizations of the release 
fractions based on measured data at different facilities or (ii) parameters from diffusion and 
effusion laws which then allow the determination of the total release efficiency, as was 
established by Kirchner et al. studying the performances of the UNILAC target ion source 
systems \cite{kirchner92}. The latter approach has been used for the present study, where the 
diffusion and effusion coefficients were mostly obtained from measurements at 
UNILAC~\cite{kirchner92,kirchner97}, CERN and Dubna~\cite{beyer03}. Because of a lack of data in 
the case of Ga and In, we used diffusion coefficients of the neighboring Ge and Sn 
elements, respectively. The FEBIAD ionization efficiencies were estimated from 
efficiencies measured at ISOLDE for rare gases~\cite{penescu10}. For the metallic elements of 
interest, an interpolation in mass gives results which are compatible with the order of 
magnitude of the efficiencies quoted by Kirchner for UNILAC (30 - 50 \%~\cite{kirchner96}).  

Figure ~\ref{fig:prod} shows the results of this comparison. The in-target yields were 
estimated from the cross-section averages as described in the previous section. The extracted 
yields are in-target yields multiplied by the diffusion, effusion and ionization efficiencies, 
and have to be compared to the experimental production rates measured at UNILAC. As in the case of 
the production cross sections, the production rates also scatter a lot. However, with the cross 
section scale factor for the low-mass region of 7.3 (figure~\ref{fig:prod}a) and 4.6 for the 
mass A=100 region (figures~\ref{fig:prod}b-e), we reach a reasonable agreement which 
seems to indicate that a reduction of the calculated cross section is also needed for this 
comparison.

We note that some of the less exotic isotopes have not been produced in ideal conditions, but experimenters set their
apparatus for a short while on these nuclei to start their experiment. As for these nuclei the beam energy was therefore
certainly not optimized, the simulation codes may have even larger deficiencies.

\subsection{Excitation function of fusion-evaporation cross sections}

As mentioned above the body of experimental data for production cross sections is quite scarce. This is even worse in terms of
excitation functions where the production cross sections are measured as a function of the energy of the incident beam. We have found one
example where sufficient data are available to make a meaningful comparison. In the Bi - Po region~\cite{sudarshan17}, a few cross sections have been 
measured as a function of the incident beam energy, however, only over a short range. In figure~\ref{fig:energy}, we compare this
excitation function to the different models used in the present work. 

Interestingly, if we exclude the CNABLA model for the two A=200 nuclei, the maximum of the calculated values is rather close for the different
models. It is difficult to say whether the experimental trend is reproduced by the model predictions. For such a statement, more data over a wider
range of energies would be needed. The figure also evidences that, in case of doubt, a slightly higher energy is more convenient to move away
from the threshold effect at low energies.

\section{Summary}

We have performed a detailed study of fusion-evaporation cross sections and production rates. 
Our first finding was that there is a rather limited number of experimental data available in the literature. In addition, these data are most likely
subject to large uncertainties keeping in mind that for most of these data no experimental error bars are given in the literature. Therefore, in order to
improve the basis for this kind of studies, experimenters need to make efforts to extract cross sections or production rates with experimental uncertainties.

We found that all codes that we tested over-estimate
the experimental production cross sections or rates with factors of 4 or more. The most reliable code is maybe the GEMINI++ evaporation code coupled with the POTFUS
fusion code, where a relatively small scale factor is needed and a relatively small scatter of the different rates or cross sections is observed once the simulated
data are scaled down. The overall overestimation of the cross sections seems to increase towards the heaviest elements. A general recommendation is to divide
predicted cross sections or rates by a factor of 5 - 10 to obtain experimental production rates in reasonable agreement with "experimental reality".


\begin{thebibliography}{10}
\bibitem{cnabla}
CNABLA is the POTFUS code coupled to the ABLA evaporation code.
\end{thebibliography}
\end{document}